\documentclass[12pt]{article}
\usepackage[utf8]{inputenc}
\usepackage[english]{babel}
\usepackage{amsmath,amsfonts,amssymb,bm}
\usepackage{graphics,epstopdf,ifpdf}
\usepackage{mathrsfs,textcomp,multicol}
\usepackage{indentfirst,fancyhdr,upgreek}
\usepackage[columnwise]{lineno}
\usepackage[bookmarksnumbered=true,bookmarksopen=true,colorlinks=true,pdfborder={0 0 1},
            urlcolor=blue,linkcolor=blue,anchorcolor=blue,citecolor=blue]{hyperref}
\usepackage{indentfirst}
\usepackage{graphicx}% Include figure filesF
\usepackage{dcolumn}% Align table columns on decimal point
\usepackage{bm}% bold math
\usepackage{lipsum}
\usepackage{slashed}
\usepackage{enumitem}
\usepackage{txfonts}
\usepackage{esint}
\usepackage{xcolor}
\usepackage{mathtools}
\usepackage{transparent}
\usepackage{upgreek}
\usepackage{tabularx}
\usepackage{csquotes}
\usepackage{geometry}
\usepackage[font=footnotesize,labelfont=bf,labelsep=period]{caption}
\usepackage{tikz}
\usepackage[compat=1.1.0]{tikz-feynman}
\tikzfeynmanset{/tikzfeynman/warn luatex=false}
\usetikzlibrary{arrows,automata}
\usetikzlibrary{matrix}
\addto\captionsenglish{}

\def\gggg{\gamma\gamma\rightarrow\gamma\gamma}

\begin{document}
\title{Two-loop QED/QCD corrections for polarized $\gggg$ process in SANCphot }

\author{ 
	\parbox{\linewidth}{\centering
          S.G.~Bondarenko$^1$\thanks{E-mail:~bondarenko@jinr.ru},
          A.~Issadykov$^{1,3}$,
          L.V.~Kalinovskaya$^2$,
          A.A.~Sapronov$^2$,
          D.~Seitova$^{3}$
}
}

\date{}

\maketitle

\vskip 0.3cm

\begin{center}
{
$^1$ \it Bogoliubov Laboratory of Theoretical Physics, JINR, Dubna, 141980 Russia\\
$^2$ \it Dzhelepov Laboratory of Nuclear Problems, JINR, Dubna, 141980 Russia \\
$^3$ \it Institute for Nuclear Physics, Ministry of Energy of the Republic of Kazakhstan, Almaty, 050032, Kazakhstan
}
\end{center}

\begin{abstract}
The new version of the {\tt SANCphot} integrator has been prepared for fast and stable numerical calculations up to two loops for polarized light-by-light scattering.
One-loop modules based on the helicity formalism with massive particles 
and two-loop modules with massless particles
inside the loops are used.
The presented study is driven by the potential of polarized photon beams to probe the high energy region. 
This study is a contribution to the research program of the CEPC project being under development in China.
\end{abstract}


\section{Introduction \label{intro}}
% intro and literature

Further studies of light-by-light (LbL) scattering 
\begin{equation}
\label{gggg}
\gamma(p_1) +\gamma(p_2) \rightarrow \\
\gamma(p_3) + \gamma(p_4),
\nonumber
\end{equation}
are mainly related to the high-luminosity $e^+e^-$ colliders 
(FCC$_{ee}$~\cite{homepagesFCCee,FCC:2018evy}, 
CEPC~\cite{homepagesCEPC,CEPCStudyGroup:2018ghi,CEPCStudyGroup:2023quu}, 
ILC~\cite{homepagesILC,ILC:2013jhg},
CLIC~\cite{homepagesCLIC,CLICdp:2018cto,Zarnecki:2020ics,Brunner:2022usy}) and optionally to the collider at low energies (Genie in China~\cite{Takahashi:2018uut}).
In both cases, 
the highly polarized initial state photons are created
through the photon production during the laser 
Compton backscattering process off the high energy electron beam.
The idea of observation of polarized $\gamma\gamma$ collisions at linear colliders via Compton backscattering was proposed by the Novosibirsk group~\cite{Ginzburg:1981ik,Ginzburg:1981vm,Ginzburg:1982yr}.
Polarized photon beams, large cross sections and high luminosity allow one
to significantly improve the accuracy of measurements at photon colliders and
thus should naturally be included into theoretical calculations and corresponding codes.
 
In 2017,  
the ATLAS and CMS Collaborations at the LHC performed the first measurement of LbL scattering in heavy-ion collisions~\cite{ATLAS:2017fur,CMS:2018erd,ATLAS:2020hii}. 
From the theoretical side,
this experiment was supported by the
codes SuperChic 4 \cite{Harland-Lang:2020veo}
and {\tt GAMMA}-UPC+MadGraph5$\_$aMC@NLO \cite{Shao:2022cly,Alwall:2014hca}
at the one-loop level.
The theoretical estimation of the cross sections comprised about two standard
deviations
below the ATLAS measured value. 
Consequently, for the modern highest experimental accuracy
of a similar experiment,
corrections with a higher than one-loop level of perturbative approximation are vitally needed.

For low-energy applications, the world's first gamma-photon collider, Genie (Qini), is set to be developed. This collider aims to conduct direct experimental tests on light-by-light scattering, focusing on achieving the maximum cross section at a center-of-mass energy of 1-2~MeV \cite{Takahashi:2018uut}. The theoretical framework will be supported by the CAIN 2.42 code ~\cite{Chen:1997fz, cain}. In the MeV energy range, it is crucial to apply a specific limit for helicity, as all amplitudes with 
an odd number of positive helicities vanish in the low-energy limit. This vanishing occurs at all-loop orders. However, for experiments like Genie, it is essential to consider the massive case. In the near future, we are planning to calculate the low-energy limit for the massive case of two-loop helicity.

The previous version of   {\tt SANCphot} integrator~\cite{Bondarenko:2022ddm} was intended for calculations of the
processes $\gamma\gamma \to \gamma\gamma (\gamma Z, ZZ)$ at the one-loop precision level.
In the {\tt SANC} project~\cite{Andonov:2004hi}, helicity formalism was always used. 
In this approach it is easy to add higher-order corrections and it is extremely convenient to estimate polarization~\cite{Gounaris:1998qk,Gounaris:1999gh}.
This paper provides a description of the new version of the {\tt SANCphot} integrator for fast and stable
numerical calculations up to two QED/QCD loops for LbL scattering.
The current version is
adjusted to study the two-loop results in the ultra-relativistic limit
(kinematic invariants $s, t, u$ 
are much greater than the
squared masses of charged fermions)
in addition to {\tt SANC} one-loop level calculations with massive particles ~\cite{Bardin:2006sn,Bardin:2009gq} 
inside the loops.
The expressions for two-loop amplitudes
in terms of fermion loops
correspond to the massless
QCD- and QED-case~\cite{Bern:2001dg}. 

The two-loop corrections to LbL scattering were computed two decades ago~\cite{Bern:2001dg,Binoth:2002xg}. In these works, the massless amplitudes were considered, as they are deemed to be most computationally accessible.
Recently, the two-loop massive QCD and QED helicity amplitudes for light-by-light scattering
were presented in \cite{AH:2023ewe,AH:2023kor}.

This article consists of four Sections.
The implementation of the two-loop QED/QCD corrections for polarized LbL cross sections
within the helicity approach is described in Section 2.
In Section 3 we present a comparison and numerical results. 
Finally, the summary is given in Section 4.

\section{Contribution of the two-loop QED/QCD corrections} 

In this section, a short explanation of the implementation of two-loop QED/QCD corrections
by using the helicity amplitudes is presented.

\begin{figure}[!h]
    \centering   \includegraphics[width=0.8\linewidth]{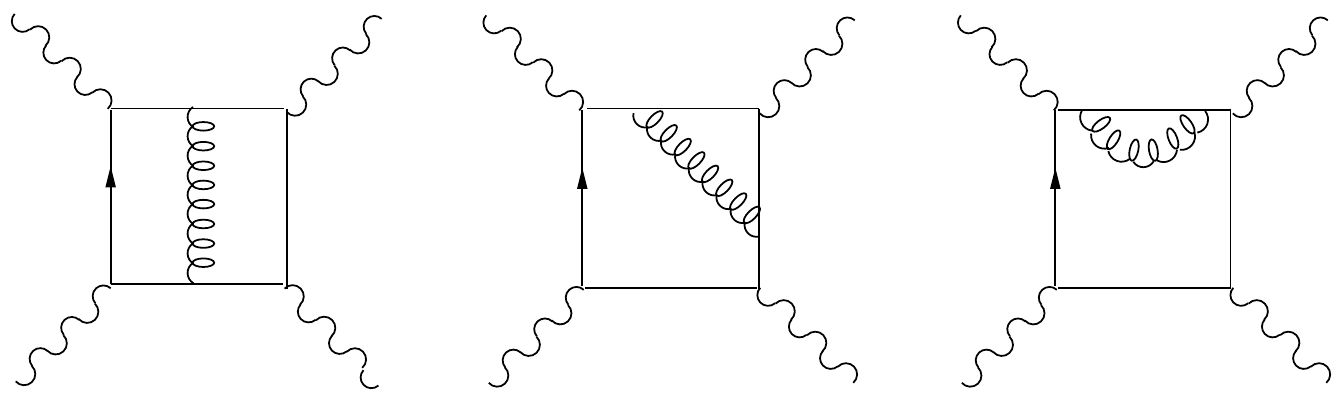}
    \caption{Two-loop diagrams corresponding to internal photon/gluon exchange}
    \label{fig:FDiag}
\end{figure}

The basic set of two-loop diagrams consists of
three types of nonvanishing
Feynman diagrams corresponding to the connection of any two sides
of the box with the photon QED or gluon QCD propagator; see Fig.~(\ref{fig:FDiag}).
The complete numbers of sixty nonvanishing Feynman diagrams are obtained by all possible permutations.
All other types of QCD diagrams vanish by Furry’s theorem or simple group theory.

The QCD- and QED-corrected two-loop amplitudes are given by the following expressions~\cite{Bern:2001dg}:
\begin{equation}
{\cal{H}}^{\text{2-loop, QCD(QED)}}_{
\lambda_1\lambda_2\lambda_3 \lambda_4}=N^{\text{QCD(QED)}} {\cal{H}}^{(2)}_{
\lambda_1\lambda_2\lambda_3 \lambda_4}
\end{equation}
with the coefficients 
$N^{\text{QCD}}= 4 (N^2 - 1)Q^4 
\alpha^2
\dfrac{\alpha_s(\mu)}{\pi}$, 
and $N^{\text{QED}}=
8 N Q^6 
\alpha^2
\dfrac{\alpha(\mu)}{\pi}$, where
$N$ is the fermion color factor (3 for quarks, 1 for leptons), and $Q$ is the fermion
charge in units of $e$.

\section{Polarized photon beams}

Taking into account parity invariance and Bose statistics, the
cross section of polarized LbL scattering in a simplified form
was derived in~\cite{Gounaris:1998qk,Gounaris:1999gh},
and we use it in our calculations~\cite{Bondarenko:2022ddm}.

The expressions for the helicity amplitudes ${\cal{H}}^{(2)}$ in the $m_f=0$ limit
(where $m_f$ is the fermion mass in a loop)  are given in~\cite{Bern:2001dg}.
Equation~(1) from~\cite{Bondarenko:2022ddm} is used
to calculate the contributions of one and two loops to the polarized cross section. 
The helicity amplitudes in the
equation are:
$${\cal H} = {\cal H}^{\text{1-loop}} + {\cal{H}}^{\text{2-loop, QCD(QED)}}.$$
In the calculations, the square of the ${\cal{H}}^{\text{2-loop}}$ amplitude
was omitted as part of the higher-order corrections.

\section{SANCphot numerical results}

The numerical results for LbL scattering  contain estimates of the total
cross sections and different
final state kinematic distributions at fixed incoming electron beam energies
\begin{equation}
\label{setEnergy}
\sqrt{s} = 250, 500, 1000 \mbox{ GeV}
\end{equation}
and various polarization setups.

The numerical validation was performed in the $\alpha(0)$ EW scheme using the setup given below:\\
\begin{tabular}{ll}
$\alpha =1/137.035990996$,      &  $\alpha_s= 0.176$,  \\
$M_{W}= 80.45149$\,GeV,           & $M_{Z}= 91.18670$\,GeV,            \\  
$M_{H}= 125$\,GeV,              &                                   \\
% $\Gamma_Z$ = 2.499770\,GeV,       & $\Gamma_W$ = 2.083600\,GeV,      \\
$m_e = 0.51099907$\,MeV,   & $m_\mu = 0.105658389$\,GeV,       \\
$m_\tau = 1.77705$\,GeV,        &                   \\
$m_u = 0.062$\,GeV,     & $m_d = 0.083$\,GeV,           \\
$m_c = 1.5$\,GeV,       & $m_s = 0.215$\,GeV,           \\
$m_t = 173.8$\,GeV,             & $m_b = 4.7 $\,GeV.           \\
\end{tabular} \\

The cross section has been integrated over a range of center-of-mass scattering angles $|\theta| > 30^{\circ}$.

To verify the technical precision of our codes, the results shown in Fig. 3  of~\cite{Bern:2001dg} for various quantities $k$ (see the definition (3.4)) were repeated.
For unpolarized initial and final photons and fully polarized ($++$ and $+-$) initial photons, 
a good agreement was observed. Qualitative agreement with the results in Fig. 4 for the QCD correction was also obtained.

In order to probe the dependence of the total cross section on the initial polarizations, 
the numerical tests were conducted using the same three polarization combinations as in~\cite{Bondarenko:2022ddm}:
\begin{eqnarray}
\label{CombPol}
 \mbox{set~1}  :&& P_e = P_e' = 0.8, \nonumber \\
&&P_{\gamma} = P_{\gamma}' = -1, P_t = P_t' = 0,
\nonumber \\
\mbox{set~2}  :&& P_e = P_e' = 0,   P_{\gamma} = P_{\gamma}' = 0,
\nonumber \\
&& P_t = P_t' = 1, \phi=\pi/2,
 \nonumber\\
 \mbox{set~3} : &&P_e = 0.8, P_e' = 0, 
P_{\gamma} = -1, P_{\gamma}' = 0, 
\nonumber \\
 &&P_t = 0, P_t' = 1, \phi=\pi/2,
\end{eqnarray}
where $P_e$, $P_\gamma$ and $P_\gamma'$ denote the helicities of the initial electron, laser and scattered photon
\cite{Kuhn:1992fx}.
We consider three cases for 
laser photons when
either both laser photons are purely transversely polarized, 
i.e.  $P_t = P_t^{'} = 1$ 
or only one of them is polarized, forward or backward.

\subsection{Cross section}
The corresponding results for the total cross section for polarized LbL are presented in Table ~\ref{tab_xs_gg} where the relative corrections $\delta$ are computed as the ratios (in percent) 
of the corresponding radiative corrections (QED/QCD) to the LO level cross section, i.e.
$\delta = \sigma^\text{NLO}/\sigma^\text{LO}-1$ where LO stands for 1-loop, and NLO -- for 2-loop
contributions.

The sign and magnitude of the integrated relative corrections corresponding to the QED and QCD two-loop effects
strongly depend on the energy and polarization combination. The QED contributions are usually only
10\% of the QCD ones, which can easily be obtained from the ratio 
$N^{\text{QED}}/N^{\text{QCD}}$.
The magnitude of the relative corrections for set 1 is smaller than for sets 2 and 3.
For center-of-mass energy  $\sqrt{s_{ee}}$ = 250 GeV the magnitude of the relative QCD corrections is approximately $0.2$ \%
for set 1 and approximately $(-0.4)$--$(-0.5)$ \% for sets 2, 3. 
For $\sqrt{s_{ee}}$ = 500 GeV the magnitude of the relative QCD corrections is $(-0.4)$ \%
for set 1 and $(-1)$ \% for sets 2,3. 
For $\sqrt{s_{ee}}$ = 1000 GeV the magnitude of the relative QCD corrections is approximately $(-0.7)$ \%
for set 1 and $(-0.9)$ \% for sets 2, 3.
Thus, these contributions are rather large under the requirements for modern experimental precision.

\begin{table}[!h]
  \caption{\label{tab_xs_gg}
%  Table 3. 
  Integrated one-loop cross sections $\sigma(\gamma\gamma)$~[fb] and the corresponding relative corrections 
  $\delta$ [\%] for the process $\gamma\gamma \rightarrow \gamma \gamma$ in different polarization setups}
  \centering
\begin{tabular}{|l|r|r|r|r|}
\hline
 $\sqrt{s_{ee}}$& & 250~GeV & 500~GeV & 1~TeV \\
\hline
set 1 & $\sigma$, fb              & 22.524(1)   & 9.445(1)     & 6.940(1) \\
      & $\delta^{\text{QED}}$,\%  & 0.017(6)    & $-0.039(3)$  & $-0.062(3)$  \\
      & $\delta^{\text{QCD}}$,\%  & 0.184(6)    & $-0.415(3)$  & $-0.662(3)$ \\
%     & $\delta^{\text{sum}}$,\%  & 0.201(6)    &              &   \\ % 1l      
%     & $\delta^{\text{sum}}$,\%  & 0.331(6)    &              &   \\ % 1l1
set 2 & $\sigma$, fb              & 24.426(1)   & 8.045(1)     &  6.861(1)   \\
      & $\delta^{\text{QED}}$,\%  & $-0.038(6)$ & $-0.099(2)$  &  $-0.087(3)$ \\
      & $\delta^{\text{QCD}}$,\%  & $-0.406(6)$ & $-1.052(2)$  &  $-0.927(2)$ \\
%     & $\delta^{\text{sum}}$,\%  & -0.444(6)   &              &   \\ % 1l      
%     & $\delta^{\text{sum}}$,\%  & -0.376(6)   &              &   \\ % 1l1      
set 3 & $\sigma$, fb              & 22.320(1)   & 8.014(1)     &  6.7967(1)   \\
      & $\delta^{\text{QED}}$,\%  & $-0.042(6)$ & $-0.103(2)$  &  $-0.084(2)$  \\
      & $\delta^{\text{QCD}}$,\%  & $-0.444(6)$ & $-1.101(2)$  &  $-0.900(2)$ \\
%     & $\delta^{\text{sum}}$,\%  & -0.486(6)   &              &   \\ % 1l      
%     & $\delta^{\text{sum}}$,\%  & -0.417(6)   &              &   \\ % 1l1      
\hline
\end{tabular}
\end{table}

\subsection{Differential distributions}

\begin{figure}[htb]
  \begin{center}
\includegraphics[width=0.315\textwidth]{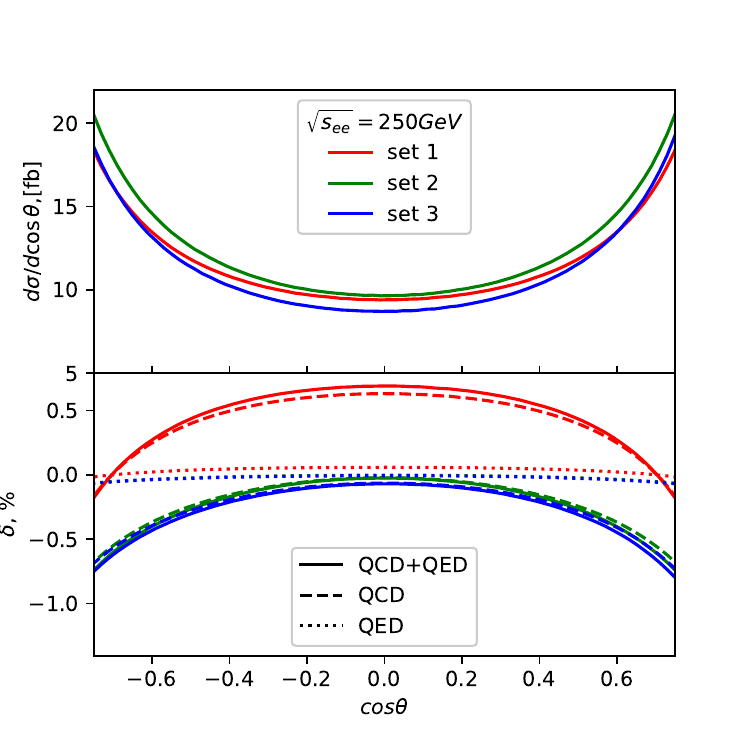}
\includegraphics[width=0.315\textwidth]{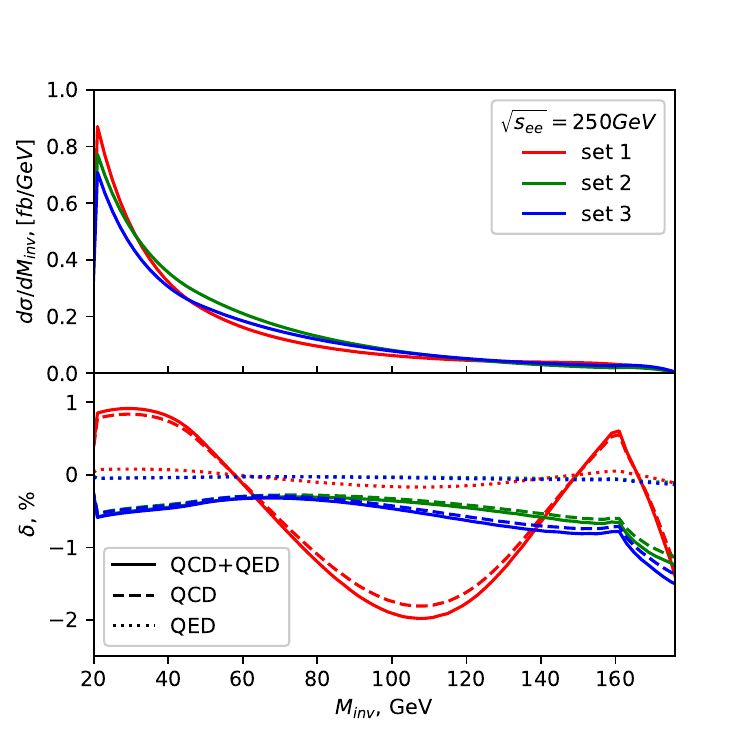}
\includegraphics[width=0.315\textwidth]{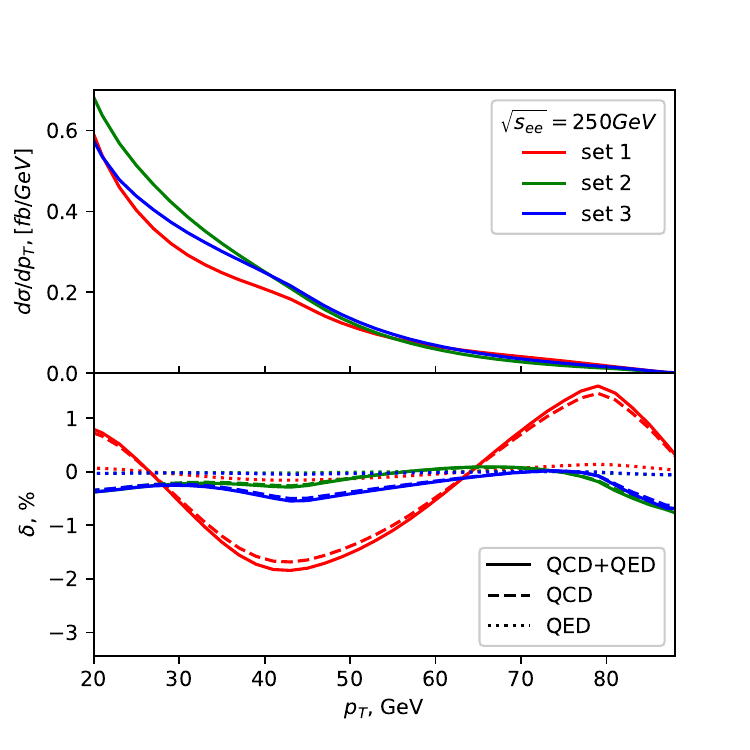} \\
\includegraphics[width=0.315\textwidth]{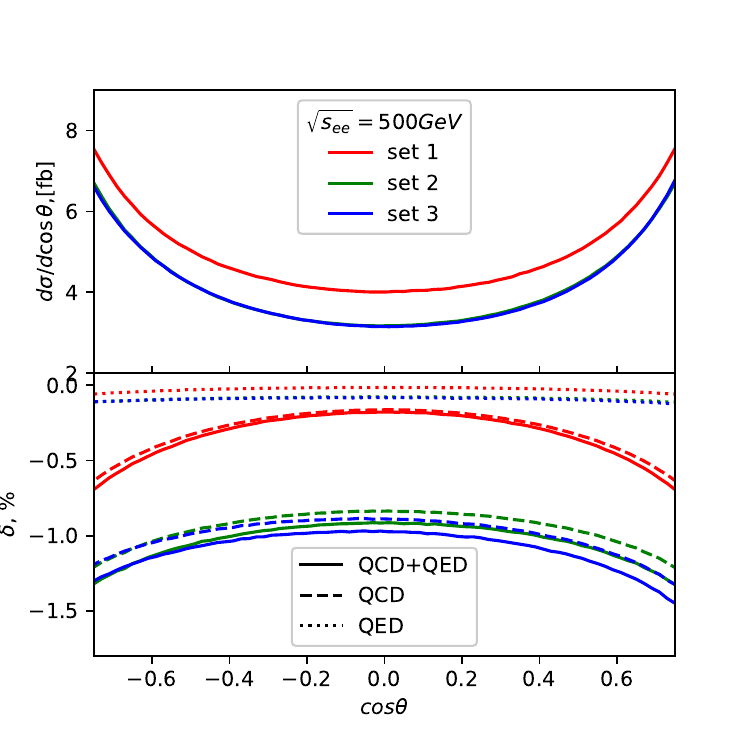}
\includegraphics[width=0.315\textwidth]{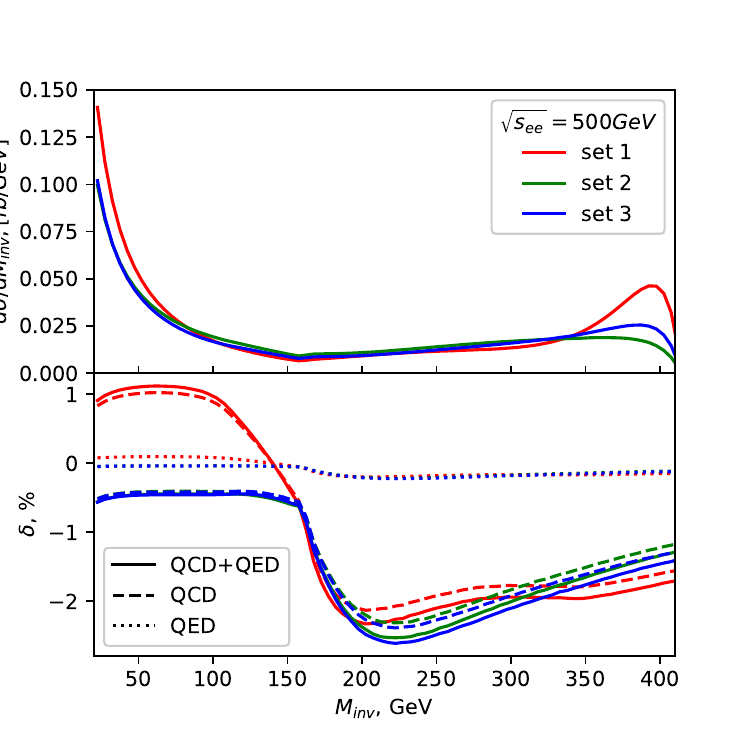}
\includegraphics[width=0.315\textwidth]{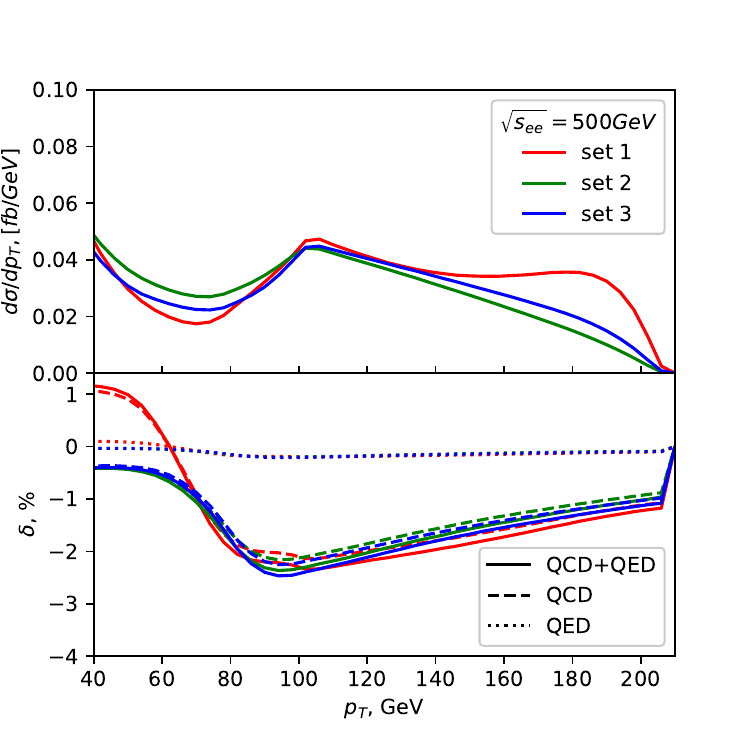} \\
\includegraphics[width=0.315\textwidth]{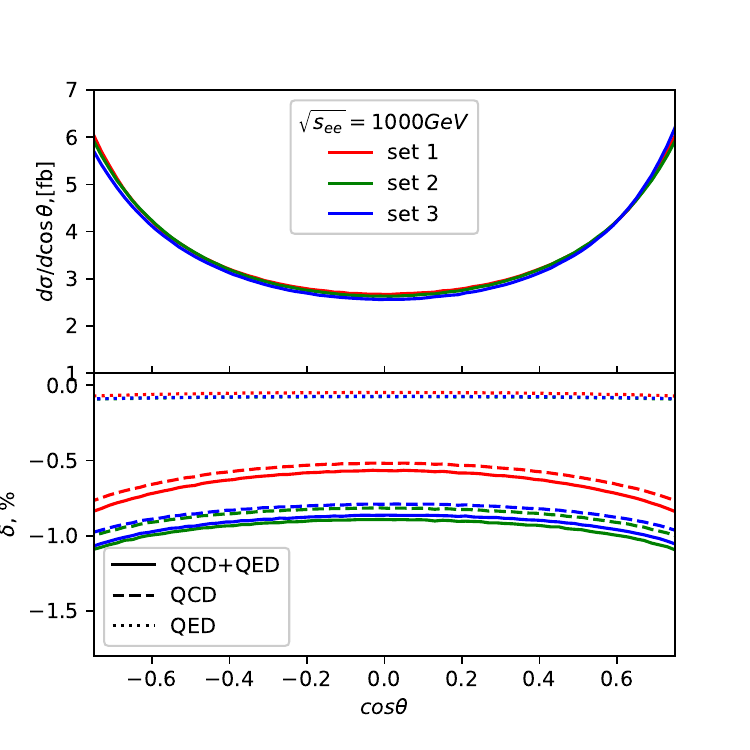}
\includegraphics[width=0.315\textwidth]{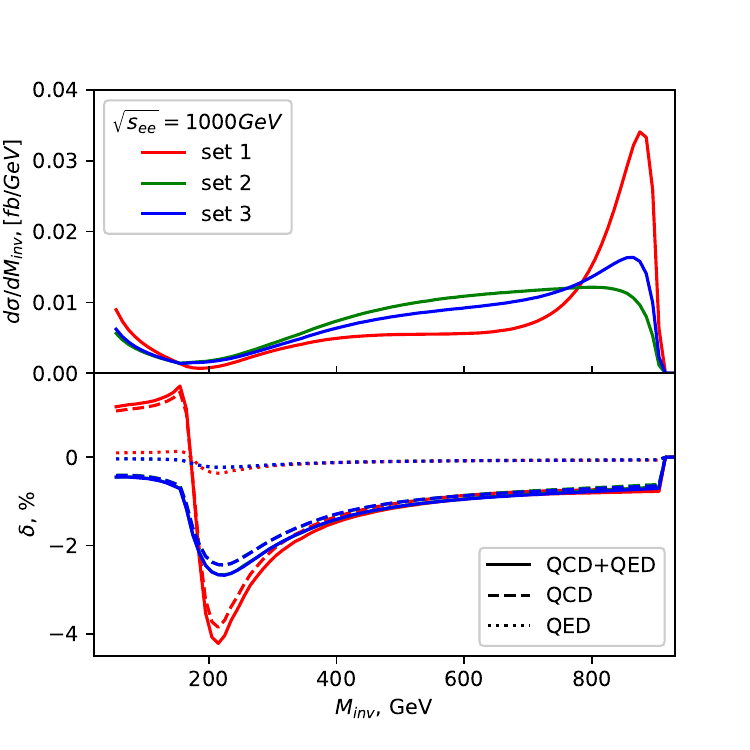}
\includegraphics[width=0.315\textwidth]{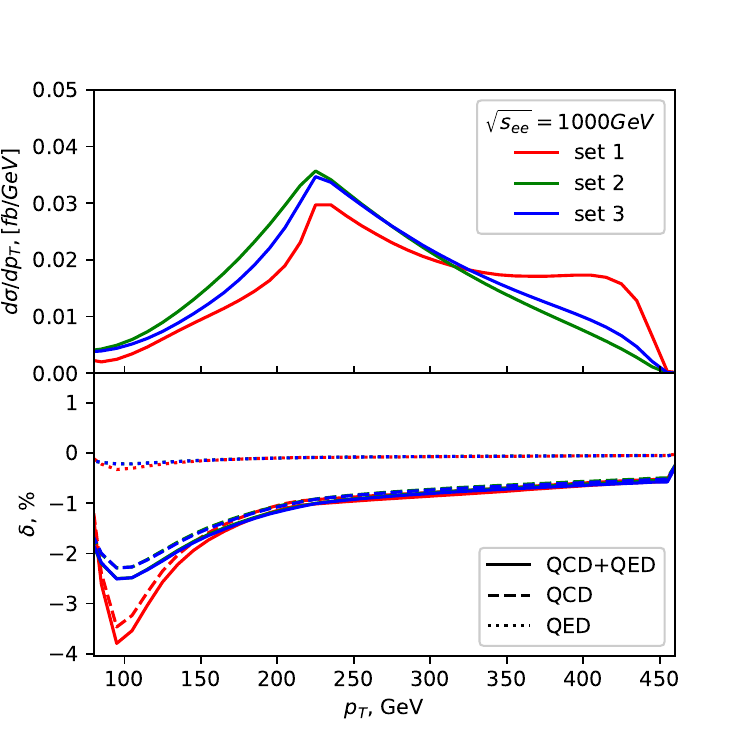} \\
  \end{center}
  \caption{\label{fig:kin_dist}
  Final state kinematic distributions of polarized LbL scattering at
  fixed incoming
   electron beam energies $\sqrt{s_{ee}} = 250, 500$ and $1000~\mathrm{GeV}$ 
  and various polarization setups~\label{fig:kin_dist_gg}}
\end{figure}

Figure~\ref{fig:kin_dist}
shows the main final-state kinematic distributions $\cos \vartheta$, $M_\text{inv}$ and $p_T$  for 
polarized LbL scattering for three different initial electron energies (\ref{setEnergy})  and
three combinations of the initial beams polarization (\ref{CombPol}).

The upper panels show the two-loop contributions of QED and QCD
to the polarized cross sections with sets~(\ref{CombPol}).
The lower panels show the relative corrections
of the QED/QCD two-loop contributions in parts.

The behavior of the relative correction distributions strongly
depends on the energy and polarization combinations. 

The results for the relative correction distributions on $\cos\theta$
for set 1 at $\sqrt{s_{ee}} = 250$ GeV are mainly positive, while the
corrections for all other energies and sets are negative. 

The relative corrections for the invariant mass $M_\text{inv}$ distributions range from $-2.5$\% to $1$\% for center-of-mass energies $\sqrt{s_{ee}}$ = 250, 500 GeV.
For  $\sqrt{s_{ee}}$ = 1000 GeV, the range widens from $-4$\% to $1.5$\%. The relative corrections for the transverse momentum $p_t$ distributions exhibit similar behavior and magnitude as those observed for the invariant mass $M_\text{inv}$ distributions. 
In summary,
the corrections for $p_t$ follow a pattern comparable to that for $M_\text{inv}$.

\section{Summary \label{sec:summary}}

The latest version of {\tt SANCphot} has been adapted to study polarized LbL
scattering. It incorporates electroweak  one-loop radiative corrections to realistic observables, as well as QCD/QED two-loop corrections that are relevant for the LbL process.  

The two-loop QED/QCD corrections make a substantial contribution. 

Including contributions up to the two-loop level, along with mass effects, allows a high degree of precision in the theoretical predictions for LbL scattering. Achieving this level of precision is essential to align with the anticipated experimental accuracy in future measurements.

\section{Funding}
\label{sec:funding}

This research has been funded by the Committee of Science of the Ministry
of Science and Higher Education of the Republic of Kazakhstan (Grant
No. AP19680084).

\label{sec:acknowledgements}

\bibliography{main.bbl}

\providecommand{\href}[2]{#2}\begingroup\begin{thebibliography}{10}

\bibitem{homepagesFCCee}
FCC-ee homepages --- {\em http://tlep.web.cern.ch}.

\bibitem{FCC:2018evy}
{FCC} Collaboration, A.~Abada {\em et al.}, {\em Eur. Phys. J. ST} {\bf 228} (2019), no.~2 261--623.

\bibitem{homepagesCEPC}
CEPC homepages --- {\em http://cepc.ihep.ac.cn/intro.html}.

\bibitem{CEPCStudyGroup:2018ghi}
{CEPC Study Group} Collaboration, M.~Dong {\em et al.}, \href{http://www.arXiv.org/abs/1811.10545}{{\tt 1811.10545}}.

\bibitem{CEPCStudyGroup:2023quu}
{CEPC Study Group} Collaboration, W.~Abdallah {\em et al.}, {\em Radiat. Detect. Technol. Methods} {\bf 8} (2024), no.~1 1--1105, \href{http://www.arXiv.org/abs/2312.14363}{{\tt 2312.14363}}.

\bibitem{homepagesILC}
ILC homepages --- {\em https://www.linearcollider.org/ILC}.

\bibitem{ILC:2013jhg}
{ILC} Collaboration, \href{http://www.arXiv.org/abs/1306.6352}{{\tt 1306.6352}}.

\bibitem{homepagesCLIC}
CLIC homepages --- {\em http://clic-study.web.cern.ch}.

\bibitem{CLICdp:2018cto}
{CLICdp, CLIC} Collaboration, T.~K. Charles {\em et al.}, \href{http://www.arXiv.org/abs/1812.06018}{{\tt 1812.06018}}.

\bibitem{Zarnecki:2020ics}
{CLICdp, ILD concept group} Collaboration, A.~F. Zarnecki, {\em PoS} {\bf CORFU2019} (2020) 037, \href{http://www.arXiv.org/abs/2004.14628}{{\tt 2004.14628}}.

\bibitem{Brunner:2022usy}
O.~Brunner {\em et al.}, \href{http://www.arXiv.org/abs/2203.09186}{{\tt 2203.09186}}.

\bibitem{Takahashi:2018uut}
T.~Takahashi {\em et al.}, {\em Eur. Phys. J. C} {\bf 78} (2018), no.~11 893, [Erratum: Eur.Phys.J.C 82, 404 (2022)], \href{http://www.arXiv.org/abs/1807.00101}{{\tt 1807.00101}}.

\bibitem{Ginzburg:1981ik}
I.~F. Ginzburg, G.~L. Kotkin, V.~G. Serbo, and V.~I. Telnov, {\em JETP Lett.} {\bf 34} (1981) 491--495.

\bibitem{Ginzburg:1981vm}
I.~F. Ginzburg, G.~L. Kotkin, V.~G. Serbo, and V.~I. Telnov, {\em Nucl. Instrum. Meth.} {\bf 205} (1983) 47--68.

\bibitem{Ginzburg:1982yr}
I.~F. Ginzburg, G.~L. Kotkin, S.~L. Panfil, V.~G. Serbo, and V.~I. Telnov, {\em Nucl. Instrum. Meth. A} {\bf 219} (1984) 5--24.

\bibitem{ATLAS:2017fur}
{ATLAS} Collaboration, M.~Aaboud {\em et al.}, {\em Nature Phys.} {\bf 13} (2017), no.~9 852--858, \href{http://www.arXiv.org/abs/1702.01625}{{\tt 1702.01625}}.

\bibitem{CMS:2018erd}
{CMS} Collaboration, A.~M. Sirunyan {\em et al.}, {\em Phys. Lett. B} {\bf 797} (2019) 134826, \href{http://www.arXiv.org/abs/1810.04602}{{\tt 1810.04602}}.

\bibitem{ATLAS:2020hii}
{ATLAS} Collaboration, G.~Aad {\em et al.}, {\em JHEP} {\bf 03} (2021) 243, [Erratum: JHEP 11, 050 (2021)], \href{http://www.arXiv.org/abs/2008.05355}{{\tt 2008.05355}}.

\bibitem{Harland-Lang:2020veo}
L.~A. Harland-Lang, M.~Tasevsky, V.~A. Khoze, and M.~G. Ryskin, {\em Eur. Phys. J. C} {\bf 80} (2020), no.~10 925, \href{http://www.arXiv.org/abs/2007.12704}{{\tt 2007.12704}}.

\bibitem{Shao:2022cly}
H.-S. Shao and D.~d'Enterria, {\em JHEP} {\bf 09} (2022) 248, \href{http://www.arXiv.org/abs/2207.03012}{{\tt 2207.03012}}.

\bibitem{Alwall:2014hca}
J.~Alwall, R.~Frederix, S.~Frixione, V.~Hirschi, F.~Maltoni, O.~Mattelaer, H.~S. Shao, T.~Stelzer, P.~Torrielli, and M.~Zaro, {\em JHEP} {\bf 07} (2014) 079, \href{http://www.arXiv.org/abs/1405.0301}{{\tt 1405.0301}}.

\bibitem{Chen:1997fz}
P.~Chen, T.~Ohgaki, A.~Spitkovsky, T.~Takahashi, and K.~Yokoya, {\em Nucl. Instrum. Meth. A} {\bf 397} (1997) 458--464, \href{http://www.arXiv.org/abs/physics/9704012}{{\tt physics/9704012}}.

\bibitem{cain}
{K. Yokoya}, {\em {User manual of CAIN, version 2.40}} (2018).

\bibitem{Bondarenko:2022ddm}
S.~G. Bondarenko, L.~V. Kalinovskaya, and A.~A. Sapronov, {\em Comput. Phys. Commun.} {\bf 294} (2024) 108929, \href{http://www.arXiv.org/abs/2201.04350}{{\tt 2201.04350}}.

\bibitem{Andonov:2004hi}
A.~Andonov, A.~Arbuzov, D.~Bardin, S.~Bondarenko, P.~Christova, L.~Kalinovskaya, G.~Nanava, and W.~von Schlippe, {\em Comput. Phys. Commun.} {\bf 174} (2006) 481--517, [Erratum: Comput.Phys.Commun. 177, 623--624 (2007)], \href{http://www.arXiv.org/abs/hep-ph/0411186}{{\tt hep-ph/0411186}}.

\bibitem{Gounaris:1998qk}
G.~J. Gounaris, P.~I. Porfyriadis, and F.~M. Renard, {\em Phys. Lett. B} {\bf 452} (1999) 76--82, [Erratum: Phys.Lett.B 513, 431--431 (2001), Erratum: Phys.Lett.B 464, 350--350 (1999)], \href{http://www.arXiv.org/abs/hep-ph/9812378}{{\tt hep-ph/9812378}}.

\bibitem{Gounaris:1999gh}
G.~J. Gounaris, P.~I. Porfyriadis, and F.~M. Renard, {\em Eur. Phys. J. C} {\bf 9} (1999) 673--686, \href{http://www.arXiv.org/abs/hep-ph/9902230}{{\tt hep-ph/9902230}}.

\bibitem{Bardin:2006sn}
D.~Bardin, L.~Kalinovskaya, V.~Kolesnikov, and E.~Uglov, ``{Light-by-light scattering in SANC}'', in {\em {International School-Workshop on Calculations for Modern and Future Colliders}}, 11, 2006, \href{http://www.arXiv.org/abs/hep-ph/0611188}{{\tt hep-ph/0611188}}.

\bibitem{Bardin:2009gq}
D.~Bardin, L.~Kalinovskaya, and E.~Uglov, {\em Phys. Atom. Nucl.} {\bf 73} (2010) 1878--1888, \href{http://www.arXiv.org/abs/0911.5634}{{\tt 0911.5634}}.

\bibitem{Bern:2001dg}
Z.~Bern, A.~De~Freitas, L.~J. Dixon, A.~Ghinculov, and H.~L. Wong, {\em JHEP} {\bf 11} (2001) 031, \href{http://www.arXiv.org/abs/hep-ph/0109079}{{\tt hep-ph/0109079}}.

\bibitem{Binoth:2002xg}
T.~Binoth, E.~W.~N. Glover, P.~Marquard, and J.~J. van~der Bij, {\em JHEP} {\bf 05} (2002) 060, \href{http://www.arXiv.org/abs/hep-ph/0202266}{{\tt hep-ph/0202266}}.

\bibitem{AH:2023ewe}
A.~A~H, E.~Chaubey, and H.-S. Shao, {\em JHEP} {\bf 03} (2024) 121, \href{http://www.arXiv.org/abs/2312.16966}{{\tt 2312.16966}}.

\bibitem{AH:2023kor}
A.~A~H, E.~Chaubey, M.~Fraaije, V.~Hirschi, and H.-S. Shao, {\em Phys. Lett. B} {\bf 851} (2024) 138555, \href{http://www.arXiv.org/abs/2312.16956}{{\tt 2312.16956}}.

\bibitem{Kuhn:1992fx}
J.~H. Kuhn, E.~Mirkes, and J.~Steegborn, {\em Z. Phys. C} {\bf 57} (1993) 615--622.

\end{thebibliography}\endgroup

\end{document}